\documentclass[aoas,preprint]{imsart}

\RequirePackage[OT1]{fontenc}
\RequirePackage{amsthm,amsmath,amssymb}
\RequirePackage{natbib}
\RequirePackage[colorlinks,citecolor=blue,urlcolor=blue]{hyperref}
\RequirePackage{float}
\RequirePackage{adjustbox}
\RequirePackage{algorithm,algorithmic} 
\RequirePackage{comment}

\startlocaldefs
\numberwithin{equation}{section}
\theoremstyle{plain}

\endlocaldefs

\begin{document}

\begin{frontmatter}
\title{Identifying main effects and interactions among exposures using Gaussian processes}
\runtitle{MixSelect using Gaussian processes}

\begin{aug}
\author{\fnms{Federico} \snm{Ferrari}\ead[label=e1]{ff31@duke.edu}}
\and
\author{\fnms{David B.} \snm{Dunson}\ead[label=e2]{dunson@duke.edu}}


\runauthor{Ferrari and Dunson}

\affiliation{Duke University}

\address{Federico Ferrari\\
Department of Statistical Science\\
Duke University\\
415 Chapel Dr, Durham, NC 27705 \\
\printead{e1}}

\address{David B. Dunson\\
Department of Statistical Science\\
Duke University\\
415 Chapel Dr, Durham, NC 27705 \\
\printead{e2}}

\end{aug}

\begin{abstract}
This article is motivated by the problem of studying the joint effect of different chemical exposures on human health outcomes.  This is essentially a nonparametric regression problem, with interest being focused not on a black box for prediction but instead on selection of main effects and interactions.  For interpretability, we decompose the expected health outcome into a linear main effect, pairwise interactions, and a nonlinear deviation.  Our interest is in model selection for these different components, accounting for uncertainty and addressing non-identifiability between the linear and nonparametric components of the semiparametric model.  We propose a Bayesian approach to inference, placing variable selection priors on the different components, and developing a Markov chain Monte Carlo (MCMC) algorithm.  A key component of our approach is the incorporation of a heredity constraint to only include interactions in the presence of main effects, effectively reducing dimensionality of the model search.  We adapt a projection approach developed in the spatial statistics literature to enforce identifiability in modeling the nonparametric component using a Gaussian process. We also employ a dimension reduction strategy to sample the nonlinear random effects that aids the mixing of the MCMC algorithm. The proposed MixSelect framework is evaluated using a simulation study, and is illustrated using data from the National Health and Nutrition Examination Survey (NHANES). Code is available on GitHub.
\end{abstract}


\end{frontmatter}

\section{Introduction}

Humans are exposed to mixtures of different chemicals arising due to environmental contamination. Certain compounds, such as heavy metals and mercury, are well known to be toxic to human health, whereas very little is known about how complex mixtures impact health outcomes. One of the key questions that epidemiology should address according to \citep{braun2016can} is: \textit{What is the interaction among agents?} The primary focus of epidemiology and toxicology studies has been on examining chemicals one at a time. However, chemicals usually co-occur in the environment or in synthetic mixtures, and hence assessing joint effects is of critical public health concern. Certainly, findings from one chemical at a time studies may be misleading  \citep{dominici2010protecting}, \citep{mauderly2008there}.

Building a flexible joint model for mixtures of chemicals is suggested by the National Research Council \citep{mauderly2010air}, \citep{vedal2011does}, \citep{national2004research}. Recently, several studies have shown relationships between complex mixtures of chemicals and health or behavior outcomes. For example, \citep{Sanders2015} reviews findings on perinatal and childhood exposures to Cadmium (Cd), Manganese (Mn) and metal mixtures.  
Several attempts have been made to simultaneously detect the effect of different chemicals on health outcomes, using either parametric or nonparametric regression techniques. The former include regularization methods, like LASSO  \citep{roberts2005critical} or Ridge Regression, and deletion/substitution/addition algorithms \citep{sinisi2004deletion}, \citep{mortimer2008air}. Some of these techniques have also been applied to high dimensional spaces  \citep{hao2014interaction}. 
While providing interpretability in terms of linear effects and pairwise interactions, the resulting dose response surface is typically too restrictive, as chemicals often have nonlinear effects. 

Nonparametric models can also be used to estimate interactions among chemicals, ranging from tree based methods \citep{hu2008temperature}, \citep{lampa2014identification}, to Bayesian Kernel Machine Regression (BKMR) \citep{BKMR2014}, \citep{valeri2017joint}, \citep{liu2017lagged} and Bayesian P-splines \citep{lang2004bayesian}. Although tree based methods, like Boosted Trees or Random Forests, are convenient computationally and often provide accurate predictions, interpretation of covariate effects is typically opaque. While providing good predictive performance, nonparametric regression surfaces like BKMR provide excessive flexibility when a simple parametric model provides an adequate approximation. On the other hand, the estimation of interactions with Bayesian P-splines becomes extremely challenging when $p$ is larger than $\sim10$, which is common in environmental epidemiology; refer to \textit{Section 2} of the \textit{\href{https://nbviewer.jupyter.org/github/fedfer/fedfer.github.io/blob/master/gp_AoAS_supplementary.pdf}{Supplementary Materials}} for additional details.

Our goal is to simultaneously estimate a flexible nonparametric model and provide interpretability. To do so, we decompose the regression surface on the health outcome into a linear effect, pairwise interactions and a nonlinear deviation. This specification, which we describe in \textit{Section 2}, allows one to interpret the parametric portion of the model while also providing flexibility via the nonparametric component. We address identifiability between the parametric and nonparametric part of the model by adapting a projection approach developed in spatial statistics, see \textit{Section 2.1}. We accurately take into account uncertainty in model selection on the different components of the model with a Bayesian approach to inference. We choose spike and slab priors for main effects and pairwise interactions \citep{george1997approaches} and allow for variable selection of nonlinear effects adapting the approach of \citep{savitsky2011}, which introduces spike and slab priors in the Gaussian process setting. We reduce computation imposing a heredity condition \citep{chipman1996bayesian}, described in \textit{Section 2.2}, and applying a dimension reduction approach to the Gaussian process surface \citep{haran2018}, \citep{banerjee2012}, which we describe in \textit{Section 3}. 

We describe our efficient Bayesian inference procedure in \textit{Section 3} and we propose a Markov chain Monte Carlo (MCMC) algorithm. We compare our method with the state of the art nonparametric models and with methods for interaction estimation in \textit{Section 4}. Finally, in \textit{Section 5} we assess the association of metal concentrations on BMI using data from the National Health and Nutrition Examination Survey (NHANES). This application shows the practical advantages of our method and how it could be used as a building block for more complex analysis.

\section{MixSelect Modeling Framework}

Let $y_i$ denote a continuous health outcome for individual $i$, let $x_i = (x_{i1},\ldots,x_{ip})^T$ denote a vector of `exposure' measurements, and let
$z_i = (z_{i1},\ldots,z_{iq})^T$ denote covariates.  For example, `exposure' may consist of the levels of different chemicals in a blood or urine sample, while covariates correspond to demographic factors and potential confounders.  For interpretability our focus is on decomposing the impact of the exposures into linear main effects, linear pairwise interactions, and a nonparametric deviation term, while including an adjustment for covariates.  Each of the exposure effect components will include a variable selection term so that some exposures may have no effect on the health response, while others only have linear main effects, and so on.  This carefully structured semiparametric model differs from usual black-box nonparametric regression analyses that can characterize flexible joint effects of the exposures but lack interpretability and may be subject to overfitting and the curse of dimensionality.  By including variable selection within our semiparametric model, we greatly enhance interpretability, while also favoring a more parsimonious representation of the regression function.

Our model structure can be described as follows: 
\begin{gather} \label{eq:model1}  
\begin{split}
    y_i & =  x_i^T\beta + \sum_{j=1}^p \sum_{k>j} \lambda_{jk}x_{ij}x_{ik} + g^*(x_i) + z_i^T\alpha + \epsilon_i,\quad \epsilon_i \sim N(0,\sigma^2), \\
g^*_n & = P g_n,\quad \quad g \sim \mbox{GP}(0,c), 
\end{split}
\end{gather}
where $\beta = (\beta_1,\ldots,\beta_p)^T$ are linear main effects of exposures, $\lambda = \{ \lambda_{jk} \}$ are pairwise linear interactions, $g_n = [g(x_1),\cdots,g(x_n)]$ is a nonparametric deviation, and $\alpha = (\alpha_1,\ldots,\alpha_q)^T$ are coefficients for the covariates.  We include variable selection in each of the three terms characterizing the exposure effects, as we will describe in detail in Section 2.2.  In addition, a key aspect of our model is the inclusion of a constraint on the nonparametric deviation to enforce identifiability separately from the linear components.  This is the reason for the $P$ term multiplying $g$ in the above expression, with $P$ a projection matrix to be described in Section 2.1.  The notation $\mbox{GP}(0,c)$ denotes a Gaussian process (GP) centered at zero with covariance function $c$ controlling the uncertainty and smoothness of the realizations.  

In spatial statistics it is common to choose a Matern covariance function, but in our setting we instead use a squared exponential covariance to favor smooth departures from linearity; in particular, we let 
\begin{align}
c(x,x') = \mbox{cov}\{ g(x),g(x') \} = \tau^2 \exp \bigg\{ \sum_{j=1}^p \rho_j (x_j-x'_j)^2 \bigg\},
\end{align}
where $\rho_j$ is a smoothness parameter specific to the $j$th exposure and $\tau^2$ is the signal variance.  Similar covariance functions are common in the machine learning literature, and are often referred to as automatic relevance determination (ARD) kernels \citep{qi2004predictive}.  They have also been employed by \citep{BKMR2014}.  However, to our knowledge  previous work has not included linear main effects and interactions or a projection adjustment for identifiability.  The proposed GP covariance structure allows variable selection ($\rho_j=0$ eliminates the $j$th exposure from the nonparametric deviation) and different smoothness of the deviations across the exposures that are included.  For example, certain exposures may have very modest deviations while others may vary substantially from linearity. 

The proposed model structure is quite convenient computationally, leading to an efficient Markov chain Monte Carlo (MCMC) algorithm, which mostly employs Gibbs sampling steps.  We will describe the details of this algorithm in Section 3, but we note that the projection adjustment for identifiability greatly aids mixing of the MCMC, and our code can be run efficiently for the numbers of exposures typically encountered in environmental epidemiology studies (up to one hundred).  Code for implementation is available at  \url{https://github.com/fedfer/MixSelect}, and also includes a modification to accommodate binary responses, as is common in epidemiology studies.  

\subsection{Non-Identifiabilty and Projection}

 Confounding between the Gaussian process prior and parametric functions is a known problem in spatial statistics and occurs when spatially dependent covariates are strongly correlated with spatial random effects, see \citep{hanks2015} or \citep{haran2018}. This problem is exacerbated when the same features are included in both the linear term and in the nonparametric surface. For this reason we project the nonlinear random effects $g$ on the orthogonal column space of the matrix containing main effects. 


The usual projection matrix on the column space of $X$ is equal to $P_X = X (X^{T} X)^{-1} X^{T}$. We define $P = P_X^\perp = I_n -P_X$ and set $g^*_n = P g_n$. Firstly, notice that the projection has an effect on the variance of the generated nonlinear effects, in particular:
    \begin{align*}
        \sum_{i = 1}^n (g^*_{i,n})^2 \le \sum_{i = 1}^n (g_{i,n})^2
    \end{align*}
    This follows from
    \begin{align*}
        (g_n^*)^T g_n^* & =  \big[ (I_n - P_X)g_n \big]^T \big[ (I_n - P_X)g_n \big]  = \\
        & = g_n^T g_n - (P_X g_n)^T (P_X g_n) \le g_n^T g_n .
    \end{align*}
\textit{Figure 1} in the \textit{\href{https://nbviewer.jupyter.org/github/fedfer/fedfer.github.io/blob/master/gp_AoAS_supplementary.pdf}{Supplementary Materials}} shows examples of realizations of $g_n$ and $g_n^*$. The curvature of the functions drawn from the projected GP is greater than the curvature in the non-projected case.

Another possibility would be to project the nonlinear random effects $g_n$ on the orthogonal column space of the matrix containing both main effects and interactions. However, we noticed in our simulations that this would make the resulting nonparametric surface too restrictive, especially when the number of possible interactions $\frac{p(p-1)}{2}$ is greater than $n$, resulting in a worse performance of the model. We did not experience significant confounding between the interaction effects and the nonlinear regression surface. 
Finally, notice that rather than sampling $g$ and then projecting onto the orthogonal column space of $X$, we can equivalently sample $g^*$ from a Gaussian process with covariance matrix $PcP^T$. Another option that we explore in \textit{Section 3} consists in integrating out the nonlinear effects.

\subsection{Variable Selection}

In this section we describe the variable selection approach that we develop in order to provide uncertainty quantification and achieve parsimonious model specification. We assume that the chemical measurements and the covariates have been standardized prior to the analysis. We choose spike and slab priors for the main effects and nonlinear effects. Regarding main effects, we choose a mixture of a normal distribution with a discrete Dirac delta at zero. Let us define as $\gamma_k$ the indicator variable that is equal to $1$ if the $k^{th}$ variable is active in the linear main effect component of the model and equal to $0$ otherwise. We have that $\beta_k \sim \gamma_k N(0, 1) +  (1-\gamma_k) \delta_0$. For the $\gamma_k$ we assume independent Bernoulli priors with success probability $\pi$. We endow $\pi$ with a Beta distribution with parameters $(a_\pi, b_\pi)$. The prior expected number of predictors included in the model is $p \frac{a_\pi}{a_\pi + b_\pi}$, which can be used to elicitate the hyperparameters $(a_\pi, b_\pi)$. As a default we choose $a_\pi = b_\pi = 1$, which corresponds to a Uniform distribution on $\pi$. We endow the main effects of covariate adjustments $\alpha_l$ with a Normal prior $N_q(0,I)$, for $l = 1, \ldots,q$. 

We impose an heredity condition for the interactions. The heredity condition is commonly employed for datasets with $p \in [20,100]$ by one-stage regularization methods like \citep{Bien2013} and \citep{Haris2018} or two stage-approaches as \citep{HaoFengZhang2016} when $p>100$. Strong heredity means that an interaction between two variables is included in the model only if the main effects are. For weak heredity it suffice to have one main effect in the model in order to estimate the interaction of the corresponding variables. Formally:
\begin{align*}
& \text{S:} \quad \lambda_{j,k} |\gamma_j =  \gamma_k = 1 \sim N(0,1),\quad \lambda_{j,k} |(\gamma_j =   \gamma_k = 1)^C \sim \delta_0  \\
& \text{W:} \quad  \lambda_{j,k} | (\gamma_j =  \gamma_k = 0)^C \sim N(0,1),\quad \lambda_{j,k} |\gamma_j =   \gamma_k = 0 \sim \delta_0 
\end{align*}
where \textit{S} and \textit{W} stand for strong and weak heredity respectively, and $\delta_0$ is a Dirac distribution at $0$. Models that satisfy the strong heredity condition are invariant to translation transformations in the covariates. Weak heredity provides greater flexibility with the cost of considering a larger number of interactions, leading to a potentially substantial statistical and computational cost. Consider the case when the $j^{th}$ covariate has a low effect on the outcome but the interaction with the $k^{th}$ feature is significantly different than zero. Strong heredity will sometimes prevent us from discovering this pairwise interaction. Heredity reduces the size of the model space from $2^{p + {p \choose 2}}$ to $\sum_{i = 0}^p {p \choose i} 2^{{i \choose 2}}$ or $ \sum_{i = 0}^p {p \choose i} 2^{p i - i (i+1)/2}$ for strong and weak heredity, respectively. The heredity condition can also be extended to higher order interactions.


As for the main effects and interactions, we apply a variable selection strategy for the nonlinear effects. 
We endow the signal standard deviation $\tau$ with a spike and slab prior, i.e. $\tau \sim \gamma^{\tau}F_\tau(\cdot)+ (1-\gamma^\tau) \delta_0$, where $F_\tau(\cdot)$ is a Gamma distribution with parameters $(1/2,1/2)$ and $\gamma^{\tau}$ has a Bernoulli(1/2) prior. We noticed that this spike and slab prior prevents overfitting of the nonlinear term in high dimensional settings, in particular when the variables are highly correlated and the true regression does not include nonlinear effects. This added benefit is highlighted in \textit{Section 4} when comparing with BKMR. Finally, when $\gamma^{\tau}  = 0$, the regression does not include nonlinear effects, resulting in faster computations. In this case, the computational complexity of the model equals the one of a Bayesian linear model with heredity constraints.

With respect to the covariate specific nonlinear effects, we follow the strategy of \citep{savitsky2011}, which is also employed by \citep{BKMR2014}, and endow the smoothness parameters $\rho_1,\cdots,\rho_p$ with independent spike and slab priors. In particular $\rho_k \sim \gamma^{\tau} \gamma_{k}^{\rho}F_\rho(\cdot)+ (1-\gamma^{\tau})(1-\gamma_{k}^\rho) \delta_0$, where $F_\rho(\cdot)$ is a Gamma distribution with parameters $(1/2,1/2)$. Only when $\gamma^{\tau}$ is different than zero, we allow the covariate specific nonlinear effects $\gamma_j^{\rho}$ to be different than zero. When $\gamma_{k}^\rho=0$, the $k^{th}$ exposure is eliminated from the nonparametric term $g$ in $(2.1)$. As before, we choose a Bernoulli prior for $\gamma^\rho_k$ with mean $\varphi$, and we endow $\varphi$ with a Beta prior with parameters $(a_\varphi,b_\varphi)$. As a default we choose $a_\varphi = b_\varphi = 1$, which corresponds to a Uniform distribution on $\varphi$. A graphical representation of the model can be found in \textit{Figure 1}.


\begin{figure}
\begin{center}
\includegraphics[width=12cm, height=10cm]{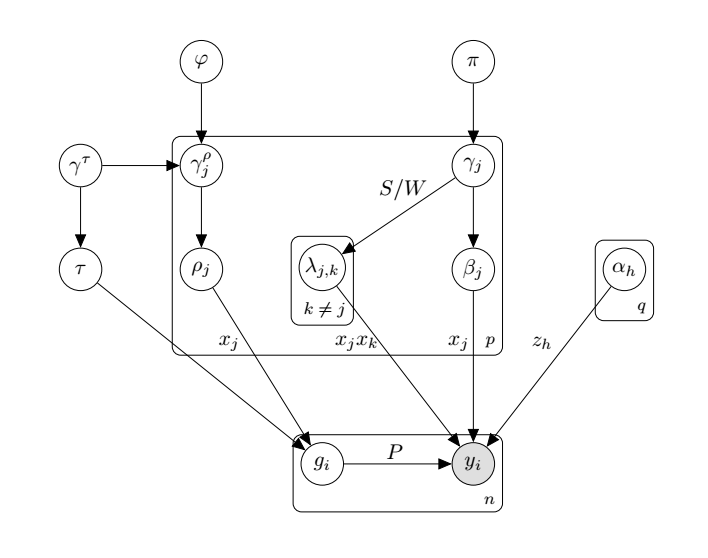}
\caption{Graphical representation of the model. The arrows between two nodes indicate conditional dependence. Variables that are in the same plate share the same indices. $S/W$ refers to strong or weak heredity.}
\end{center}
\end{figure}

\section{Computational Challenges and Inference}

In this section we describe how we conduct inference for model $(2.1)$. We also address the computational challenges associated with Gaussian process regression in the Bayesian framework and summarize the MCMC algorithm at the end of the section. 

We defined a mixture of Normal priors for the main effects, interactions and the coefficients of the covariate adjustments, namely $\beta$, $\lambda$ and $\alpha$, in \textit{Section 2.2}. Having a Gaussian likelihood, the full conditionals for these parameters are conjugate, hence we can directly sample from multivariate normal distributions within a Gibbs sampler. This operation could be quite expensive since the number of parameters is of order $p^2$. However, thanks to the strong heredity condition, we only need to sample the interactions between the variables with non-zero main effects and we set to zero all the others. 
Given each of the elements of $\beta$, $\lambda$ and $\alpha$ we can update the labels $\gamma$ with a Bernoulli draw. We also re-parametrize the model setting $\tau = \tau^* \sigma$, so that we can directly update $\sigma^2$ from an inverse Gamma distribution.

Dealing with the nonlinear term $g$ can also be expensive since we need to sample $n$ parameters at each iteration. For this reason, we integrate out the GP term so that marginally the likelihood of model $(2.1)$ is equivalent to:
\begin{align}
y | \beta,\Lambda, c \sim N(X \beta + diag(X \Lambda X^T) + \alpha Z, \sigma^2 I_n + P c P^T ),
\end{align}
where $\Lambda$ is a upper triangular matrix such that $\Lambda_{j,k} = \lambda_{j,k}$ when $k > j$ and zero otherwise. 

The covariance matrix depends on the hyperparameters $\rho_j$, for $j = 1, \ldots,p$, that define the variable selection scheme for the nonlinear effects. The priors for the smoothness parameters $\rho_j$ and $\tau^2$ defined in \textit{Section 2.2} are not conjugate so that we need a Metropolis-Hastings step within the Gibbs sampler to sample these parameters.  In order to compute the acceptance ratio, we need to evaluate the likelihood of $(2)$ and invert the matrix $\sigma^2 I_n + P c P^T$ of dimension $n$: such operation is of complexity $O(n^3)$ and needs to be done $p$ times. For this reason we approximate the matrix $ P c P^T$ with the strategy described in \textit{Algorithm 1} of \citep{haran2018}. This approach is a generalization of \citep{banerjee2012} and uses random projections to find an approximation of the Eigen Decomposition of $ P c P^T$. In particular we approximate this matrix as $U_m D_m U_m^T$, where $m$ is related to the order of the approximation, with $m$ usually being much smaller than $n$. $D_m$ is a diagonal matrix of dimension $m$ and $U_m$ is of dimension $n \times m$. We can now apply the Sherman-Morrison-Woodbury formula to compute the inverse of $\Sigma = \sigma^2 I_n + P c P^T$:
\begin{align*}
\Sigma^{-1} & = (\sigma^2 I_n + P c P^T)^{-1} \approx (\sigma^2 I_n + U_m D_m U_m^T)^{-1}  = \\ 
& = \frac{1}{\sigma^2}(I_n + U_m (\sigma^2 D_m + U_m^T U_m)^{-1}U_m^T) ,
\end{align*}
which now involves the inversion of an $m \times m$ matrix. Similarly we can simplify the computations for the determinant of $\Sigma$ using the Determinant Lemma \citep{harville1998}: 
\begin{align*}
|\Sigma|=|\sigma^2 I_n + P c P^T| \approx \sigma^{2n }  
\prod_{j = 1}^m (D_{m; j,j}^{-1} + \sigma^{-2}) D_{m; j,j} .
\end{align*}
It is challenging to design a sampler with satisfactory mixing for the smoothness parameters $\{ \rho_j \}$. However we obtained good performance for an add-delete sampler, which updates $\rho_j$ at every iteration. When the previous $\rho_j = 0$, we perform \textit{add move}: sample from a distribution with support on $\mathbb{R_+}$. When $\rho_j \neq 0$, we perform a \textit{delete move} and propose $\rho_j = 0$. Then, for the $\rho_j \neq 0$, we also perform the \textit{Gibbs-type} move and sample from the same proposal as in the \textit{add move}. The MCMC sampler is summarized in \textit{Algorithm 1}.


\section{Simulations}

In this section we compare the performance of our model with respect to five other methods: BKMR \citep{BKMR2014}, Family \citep{Haris2018}, hierNet \citep{Bien2013}, PIE \citep{wang2019penalized} and RAMP \citep{HaoFengZhang2016}. BKMR is a nonparametric Bayesian method that employs Gaussian process regression with variable selection in a similar fashion as model $(2.1)$. Family, hierNet, PIE and RAMP are designed for interaction selection in moderate to high dimensional settings. We generate the covariates independently $X_i \sim N_p(0,I_p)$  for $i = 1,\cdots,n$, for $n = 250,500$ and $p = 25,50$, so that the number of parameters that we estimate with model $(2.1)$ is $353$ and $1352$, respectively. We generate the outcome as follows:
\begin{align*}
    & \text{(a)} \quad y_i  = x_1 - x_2 + x_3 + 2 x_1 x_2 - x_1 x_3 + \frac{1}{2} x_4^2 + \frac{4}{exp(-2 x_5) + 1} +  \epsilon_i \\ 
& \text{(b)} \quad y_i  =  x_1 + x_2 - x_3 - x_4 + 2 x_1 x_2 - x_1 x_3 - x_2 x_3 -2 x_3 x_4 + \epsilon_i \\
 & \text{(c)} \quad y_i  = sin(x_1 + 3 x_3) - \frac{1}{2} x_3^2 + exp(-0.1*x_1) + \epsilon_i 
\end{align*}
where $\epsilon_i \sim N(0,1)$. The first setting involves a model with strong heredity and nonlinear effects, whereas the second is an interaction model and the third a nonlinear model. We evaluate the performance on a test dataset of $100$ units with predictive mean squared error for all the models. We compute the Frobenious norm for the matrix containing pairwise interactions for Family, hierNet, RAMP and PIE. The Frobenious norm between two square matrices $\Lambda$ and $\hat{\Lambda}$ of dimension $p$ is defined as 
\begin{align*}
\sqrt{trace((\Lambda - \hat{\Lambda})^T (\Lambda - \hat{\Lambda}))} .
\end{align*}
We also compute posterior inclusion probabilities of nonlinear effects, so that we can calculate the percentage of True positive and True negative nonlinear effects for our method and BKMR. We average the results across $50$ simulations. The results for $n = 500$ are summarized in \textit{Table 1} and  \textit{Table 2} and for $n = 250$ are summarized in \textit{Table 1} and  \textit{Table 2} of the \textit{\href{https://nbviewer.jupyter.org/github/fedfer/fedfer.github.io/blob/master/gp_AoAS_supplementary.pdf}{Supplementary Materials}}.

Across all the simulation scenarios, our model consistently achieves nearly the best predictive performance in terms of prediction error and Frobenious norm, and is able to identify main effects, interactions and nonlinear effects. The experiments highlight the advantages of MixSelect in the context of the application, where the dose-response surfaces usually have roughly linear, hill-shaped or sigmoid shapes. Hence constraining the flexible nonparametric surface allows MixSelect to have a predictive and inference advantage over BKMR, which is the main nonparametric method used in environmental epidemiology applications. For model (a), we achieve a better performance because of the decomposition of the regression surface, and we correctly identify linear and nonlinear effects. With respect to model (b), our method is able to correctly estimate a regression surface without nonlinear effects, thanks to the spike and slab prior on the term $\tau$. We also achieve a similar if not better performance in the nonlinear scenario of method (c). Finally, \textit{Figure 2} of the \textit{\href{https://nbviewer.jupyter.org/github/fedfer/fedfer.github.io/blob/master/gp_AoAS_supplementary.pdf}{Supplementary Materials}}  shows the estimated regression surface versus the true surface for model (a), when $n = 250$ and $p = 25$.

\begin{table}[p] \centering 
  \label{} 
\begin{tabular}{@{\extracolsep{5pt}} c | ccccccc} 
\\[-1.8ex]\hline 
\hline \\[-1.8ex] 
 & & MixSelect  & BKMR  & hierNet  & Family  & PIE  & RAMP  \\ 
 \hline \\[-1.8ex] 
& test MSE & $1.138$ & $1$ & $1.098$ & $5.645$ & $4.400$ & $1.217$ \\ 
& FR & $1.033$ &   & $5.659$ & $5.820$ & $2.465$ & $1$ \\ 
\cline{2-8}
model (a) & TP main & $1$ &   & $1$ & $1$ & $1$ & $1$ \\ 
& TN main & $0.758$ &   & $0.798$ & $0.947$ & $0.679$ & $0.919$ \\ 
& TP int & $1$ &   & $1$ & $1$ & $1$ & $1$ \\ 
& TN int & $1.000$ &   & $0.989$ & $0.984$ & $0.997$ & $0.997$ \\ 
& TP nl & $0.947$ & $1$ &   &   &   &   \\ 
& TN nl & $0.977$ & $0.821$ &   &   &   &   \\ 
\hline \\[-1.8ex] 
\hline \\[-1.8ex] 
& test MSE & $1$ & $1.902$ & $1.430$ & $8.928$ & $1.363$ & $1.061$ \\ 
& FR & $1$ &   & $18.162$ & $22.572$ & $1.723$ & $1.433$ \\ 
\cline{2-8}
model (b) & TP main & $1$ &   & $1$ & $1$ & $1$ & $1$ \\ 
& TN main & $0.998$ &   & $0.863$ & $0.907$ & $0.688$ & $0.992$ \\ 
& TP int & $1$ &   & $1$ & $0.978$ & $1$ & $0.989$ \\ 
& TN int & $1$ &   & $0.988$ & $0.958$ & $0.993$ & $0.999$ \\ 
& TP nl &   &   &   &   &   &   \\ 
& TN nl & $0.984$ & $0.673$ &   &   &   &   \\ 
\hline \\[-1.8ex] 
\hline \\[-1.8ex] 
& test MSE & $1.359$ & $1$ & $1.203$ & $2.927$ & $1.285$ & $2.641$ \\ 
& FR & $1$ &   & $8.759$ & $2.508$ & $9.600$ & $5.542$ \\ 
\cline{2-8}
model (c) & TP main &   &   &   &   &   &   \\ 
& TN main & $0.808$ &   & $0.719$ & $0.868$ & $0.834$ & $0.851$ \\ 
& TN int & $1.000$ &   & $0.984$ & $0.980$ & $0.992$ & $0.991$ \\ 
& TP nl & $0.645$ & $0.985$ &   &   &   &   \\ 
& TN nl & $0.989$ & $0.893$ &   &   &   &   \\ 
\hline \\[-1.8ex] 
\hline \\[-1.8ex] 
\end{tabular} 
\caption{Results from the simulation study under the three scenarios with $p = 25$, $n = 500$. We computed test error, FR for interaction effects, percentage of true positives and true negatives for main effects and interactions for MixSelect, BKMR, hierNet, Family, PIE and RAMP. We divided each value of test error and FR by the best (lowest) result for that metric. This makes the metric of the best model equal to $1$.}
\end{table}

\begin{table}[p] \centering 
  \label{} 
\begin{tabular}{@{\extracolsep{5pt}} c | ccccccc} 
\\[-1.8ex]\hline 
\hline \\[-1.8ex] 
 & & MixSelect  & BKMR  & hierNet  & Family  & PIE  & RAMP  \\ 
\hline \\[-1.8ex] 
& test MSE & $1.135$ & $11.409$ & $1$ & $5.630$ & $4.057$ & $1.181$ \\ 
& FR & $1.808$ &   & $8.718$ & $9.642$ & $3.949$ & $1$ \\ 
\cline{2-8}
model (a) & TP main & $1$ &   & $1$ & $0.993$ & $1$ & $1$ \\ 
& TN main & $0.863$ &   & $0.868$ & $0.976$ & $0.789$ & $0.967$ \\ 
& TP int & $1$ &   & $1$ & $0.989$ & $1$ & $1$ \\ 
& TN int & $1$ &   & $0.996$ & $0.996$ & $0.999$ & $1.000$ \\ 
& TP nl & $0.826$ & $1$ &   &   &   &   \\ 
& TN nl & $0.999$ & $0.037$ &   &   &   &   \\ 
\hline \\[-1.8ex] 
\hline \\[-1.8ex] 
& test MSE & $1.000$ & $12.987$ & $1.420$ & $9.485$ & $1.364$ & $1$ \\ 
& FR & $1.222$ &   & $20.973$ & $25.820$ & $1.849$ & $1$ \\ 
\cline{2-8}
model (b) & TP main & $1$ &   & $1$ & $1$ & $1$ & $1$ \\ 
& TN main & $0.999$ &   & $0.880$ & $0.977$ & $0.822$ & $0.999$ \\ 
& TP int & $1$ &   & $1$ & $0.990$ & $0.995$ & $1$ \\ 
& TN int & $1$ &   & $0.996$ & $0.993$ & $0.999$ & $1.000$ \\
& TN nl & $1$ & $0.046$ &   &   &   &   \\ 
\hline \\[-1.8ex] 
\hline \\[-1.8ex] 
& test MSE & $1.360$ & $4.139$ & $1$ & $2.589$ & $1.070$ & $2.519$ \\ 
& FR & $1$ &   & $7.990$ & $2.078$ & $8.885$ & $3.562$ \\ 
\cline{2-8}
model (c) & TN main & $0.894$ &   & $0.815$ & $0.950$ & $0.901$ & $0.942$ \\
& TP int &   &   &   &   &   &   \\ 
& TN int & $1.000$ &   & $0.994$ & $0.997$ & $0.998$ & $0.999$ \\ 
& TP nl & $0.523$ & $0.983$ &   &   &   &   \\ 
& TN nl & $0.984$ & $0.043$ &   &   &   &   \\ 
\hline \\[-1.8ex] 
\hline \\[-1.8ex] 
\end{tabular} 
\caption{Results from the simulation study under the three scenarios with $p = 50$, $n = 500$. We computed test error, FR for interaction effects, percentage of true positives and true negatives for main effects and interactions for MixSelect, BKMR, hierNet, Family, PIE and RAMP. We divided each value of test error and FR by the best (lowest) result for that metric. This makes the metric of the best model equal to $1$.}
\end{table}

\section{Environmental Epidemiology Application}

\subsection{Motivation}

The goal of our analysis is to assess the association of fourteen metals (Barium, Cadmium, Cobalt, Caesium, Molybdenum, Manganese, Mercury, Lead, Antimony, Tin, Strontium, Thallium, Tungsten and Uranium) with body mass index (BMI). Recently, several studies showed the relation between complex mixtures of metals and health or behavioral outcomes. See \citep{Sanders2015} for example for a literature review on perinatal and childhood exposures to Cadmium (Cd), Manganese (Mn) and metal mixtures. The authors state that there is suggestive evidence that Cadmium is associated with poorer cognition. \citep{ClausHenn2014} report associations between mixtures and pediatric health outcomes, cognition, reproductive hormone levels and neurodevelopment. With respect to obesity indices, metals have already been associated with an increase in waist circumference and BMI, see \citep{padilla2010examination} and \citep{shao2017association}, using data from the National Health and Nutrition Examination Survey (NHANES).

\subsection{Data Description}

We consider data from NHANES collected in 2015. We select a subsample of $2532$ individuals for which at least one measurement of metals and BMI have been recorded. We also include in the analysis cholesterol, creatinine, sex, age and ethnicity, which has five categories (Hispanic, other Hispanic, non-Hispanic White, non-Hispanic Black and other Etnicity). We choose Hispanic as a reference group for ethnicity. \textit{Table 3} in the \textit{\href{https://nbviewer.jupyter.org/github/fedfer/fedfer.github.io/blob/master/gp_AoAS_supplementary.pdf}{Supplementary Materials}} shows the correlations among chemicals, \textit{Figure 3} and \textit{Figure 4} in the \textit{\href{https://nbviewer.jupyter.org/github/fedfer/fedfer.github.io/blob/master/gp_AoAS_supplementary.pdf}{Supplementary Materials}} show the missingness pattern and the cases below the limit of detection (LOD). In NHANES, different groups of chemicals, such as metals or Phthalates, are only measured for a subsample of individuals.  This subsampling only depends on demographic characteristics of the individuals and hence the missing at random assumption should be appropriate in our context.

We apply the base $10$ logarithm transformation to the chemical exposure values, cholesterol and creatinine. We also apply the $\text{log}_{10}$ transformation to BMI in order to make its distribution closer to normality, which is the assumed marginal distribution in our model. The log-transformation is commonly applied in environmental epidemiology in order to reduce the influence of outliers and has been employed in several studies using NHANES data \citep{nagelkerke2006body}, \citep{lynch2010objectively}, \citep{buman2013reallocating}. We leave these transformations implicit for the remainder of the section.

\subsection{Missing data and LOD}

In this subsection we describe how to explicitly model the covariates to allow imputation of observations that are missing or below the limit of detection. We are particularly motivated by studies of environmental health collecting data on mixtures of chemical exposures. These exposures can be moderately high-dimensional with high correlations within blocks of variables. For this reason we decide to endow the chemical measurements, cholesterol and creatinine with a latent factor model.
Let $X$ be the $n\times p$ matrix containing the chemical measurements, $Z$ an $n\times q$ matrix containing the covariates and let $W_i=(X_i,z_{i1},z_{i2})^T$ be a $d \times 1$ vector containing the $14$ chemical measurements, cholesterol and creatinine. The factor model is as follows:
\begin{align}
& W_i = \Lambda \eta_i+ \epsilon_i,   \quad  \epsilon_i \sim N_p(0,\Sigma), \\
& \eta_i \sim N_k(0,I), \nonumber
\end{align}
where we center the data $W_i$ to have zero mean prior to the analysis, $\Sigma = diag(\sigma_1^2,\ldots, \sigma_d^2)$ is as residual variance matrix, $\Lambda$ is a $d \times k$ factor loadings matrix, and $\eta_i$ are i.i.d. standard normal latent  factors. We assume an element wise standard normal prior for $\Lambda$ and endow $\sigma_j^2$ with independent Inverse-Gamma priors with parameters $(1/2,1/2)$, for $j = 1,\ldots,d$. From an eigendecomposition of the correlation matrix, the first $9$ eigenvectors explain more than $85\%$ of the total variability; hence we set the number of factors equal to $9$.  \textit{Algorithm 2} in the \textit{\href{https://nbviewer.jupyter.org/github/fedfer/fedfer.github.io/blob/master/gp_AoAS_supplementary.pdf}{Supplementary Materials}} describes how to sample the parameters of $(5.1)$ within an MCMC algorithm. 

In addition to missingness due to chemicals that have not  been assayed, $13.5\%$ of chemicals have been recorded under the limit of detection (LOD). We can impute these observations as:
\begin{align*}
    X_{ij} | X_{ij} \in [-\infty, \text{log}_{10}(\text{LOD}_j)] \sim TN(\eta_i ^T \lambda_j, \sigma_j^2, -\infty, \text{log}_{10}(\text{LOD}_j)),
\end{align*}
where $\text{LOD}_j$ is the limit of detection for exposure $j$ and $TN(\mu,\sigma^2, a, b)$ is a truncated normal distribution with mean $\mu$, variance $\sigma^2$ and support in $[a,b]$. A related approach was used in \cite{ferrari2020bayesian} to impute chemicals below the LOD within an MCMC algorithm.

To simplify data imputation under the above model and improve robustness to model misspecification, we apply a common ``cut of feedback" approach \citep{lunn2009combining}. In particular, in imputing the missing values and those below the limit of detection, we use the conditional posterior given only the data in the $W_i$ component of the model and not taking into account that $W_i$ also appears in the outcome model.

\subsection{Statistical Analysis}

We estimate a quadratic regression with nonlinear effects for the transformed chemicals, which are included in the matrix $X$, and we control for covariates, which are included in the matrix $Z$, according to model $(2.1)$. We use the specified priors in \textit{Section 2.2} and alternate between the steps of \textit{Algorithm 1} and \textit{Algorithm 2} at each MCMC iteration to obtain the posterior samples. In environmental epidemiology, the signal to noise ratio is usually low; hence we use the weak heredity specification in order to have greater flexibility in our model and to enhance power in discovery of linear interactions. We run the MCMC chain for a total of $5000$ iterations, with a burn-in of $4000$.

We observed good mixing for main effect and interaction coefficients. In particular, the average Effective Sample Size (ESS) for main effects and interactions was equal to $725$. For the smoothness parameters, the effective sample size for each $\rho_j$ was on average $3$ times higher with respect to the corresponding parameters in BKMR. We also computed the Geweke diagnostic for main and interaction effects, for a total of $105$ parameters. The Geweke diagnostic tests for a difference of the mean in the first $25\%$ of the MCMC samples and the last $25\%$ of the samples. All computed p-values were not significant at the $0.01$ level. Residual plots are included in \textit{Figure 5} of the \textit{\href{https://nbviewer.jupyter.org/github/fedfer/fedfer.github.io/blob/master/gp_AoAS_supplementary.pdf}{Supplementary Materials}}. The residual diagnostics suggest that the model assumptions are satisfied fairly well. Firstly, approximate normality holds, with only a mild deviation in the tails. Secondly, inspecting the scatter plot of predicted BMI vs standardized residual, we did not find any clear patterns, suggesting homoskedasticity and adequate fit of our regression model. Lastly, we conducted posterior predictive checks, comparing the mean of the in-sample predictions at each MCMC iteration to the data mean. \textit{Figure 6} in the \textit{\href{https://nbviewer.jupyter.org/github/fedfer/fedfer.github.io/blob/master/gp_AoAS_supplementary.pdf}{Supplementary Materials}} shows that the two means align very well. We also observed good in sample and out of sample coverage of $100(1-\alpha)\%$ predictive intervals for different $\alpha$ values; refer to \textit{Table 4} in \textit{\href{https://nbviewer.jupyter.org/github/fedfer/fedfer.github.io/blob/master/gp_AoAS_supplementary.pdf}{Supplementary Materials}}. 

The complexity per iteration of Gibbs sampling is $\mathcal{ O}(n^2 m)$ when $\tau \neq 0$, where $m$ is related to the approximation described in \textit{Section 3}. When $\tau = 0$, the complexity per iteration of Gibbs sampling is $\mathcal{ O}(d^2)$, where $d$ is the number of active main effects.

\subsection{Results}

In our analysis, we found significant nonlinear associations with BMI for Cadmium and Tungsten with posterior predictive probabilities of having an active nonlinear effect of $1$ and $0.79$, respectively. \textit{Figure 2} shows the estimated nonlinear surfaces for Cadmium and Tungsten, when all the other variables are set to their median. The nonlinear effect of Cadmium has a hill-shaped dose response, with a monotone increase at lower doses followed by a downturn leading to a reverse in the direction of association; presumably as toxic effects at high doses lead to weigh loss. We also found a significant negative linear association between BMI and Lead and Molybdenum, and the main effect estimates suggested a negative linear association with Cesium, Cobalt and Tin. A similar negative effect for higher doses of Cadmium, Cobalt and Lead was found in \citep{shao2017association} and \citep{padilla2010examination}, where both authors found an inverse linear association between these metals and BMI, suggesting that they can create a disturbance of metabolic processes. 

We found positive linear interactions between Molybdenum$\times$Strontium, Lead$\times$Antimony, and negative interaction between Lead$\times$Uranium. \textit{Figure 7} in the \textit{\href{https://nbviewer.jupyter.org/github/fedfer/fedfer.github.io/blob/master/gp_AoAS_supplementary.pdf}{Supplementary Materials}} shows the estimated coefficients for interactions. With respect to covariate adjustments, we found a positive association between BMI and Age, Creatinine and Cholesterol, as expected, and also a negative association with ethnicities Other Hispanic, non-Hispanic White, non-Hispanic Black and Other Ethnicity with respect to the reference group Hispanic, refer to \textit{Figure 8} of the \textit{\href{https://nbviewer.jupyter.org/github/fedfer/fedfer.github.io/blob/master/gp_AoAS_supplementary.pdf}{Supplementary Materials}}. Finally, even if some of the chemicals were moderately correlated, see Molybdenum and Tungsten for example in \textit{Table 3} in the \textit{\href{https://nbviewer.jupyter.org/github/fedfer/fedfer.github.io/blob/master/gp_AoAS_supplementary.pdf}{Supplementary Materials}}, our model was able to distinguish the two effects, estimating a linear association for Molybdenum and no association for Tungsten.

We compared the performance of our model with the methods described in \textit{Section 4}: BKMR \citep{BKMR2014}, Family \citep{Haris2018}, hierNet \citep{Bien2013}, PIE \citep{wang2019penalized} and RAMP \citep{HaoFengZhang2016}. For simplicity in making comparisons across methods that mostly lack an approach to accommodate missing exposures, we focus on complete case analyses, discarding all observations having any values that are missing. \textit{Table 3} shows the performance of the models for in sample MSE when training on the full dataset and out of sample MSE when holding out $500$ data points. Notice that BKMR overfits the training data in the presence of highly correlated covariates and consequently has worse performance on the test set. In addition, BKMR estimates a posterior probability of a nonlinear effect greater than 0.87 for each chemical, which could be a result of overfitting. On the other hand, MixSelect is able to distinguish a simple regression surface from a more complex one thanks to the identifiability constraint, which prevents overfitting.

\begin{figure}[H]
\begin{minipage}{0.48\textwidth}
\centering
\includegraphics[width=\linewidth, height= 5.8cm]{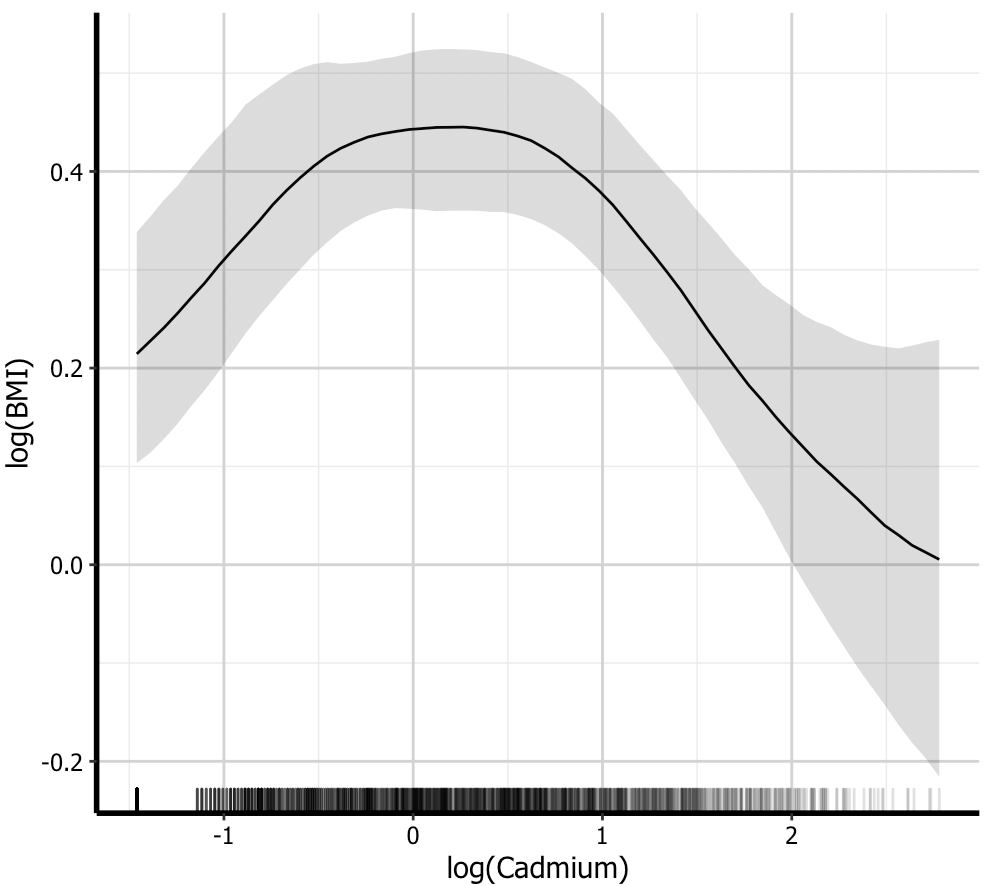}
\end{minipage} \hfill
\vspace{1.5cm}
\begin{minipage}{0.48\textwidth}
\centering
\includegraphics[width=\linewidth, height= 5.8cm]{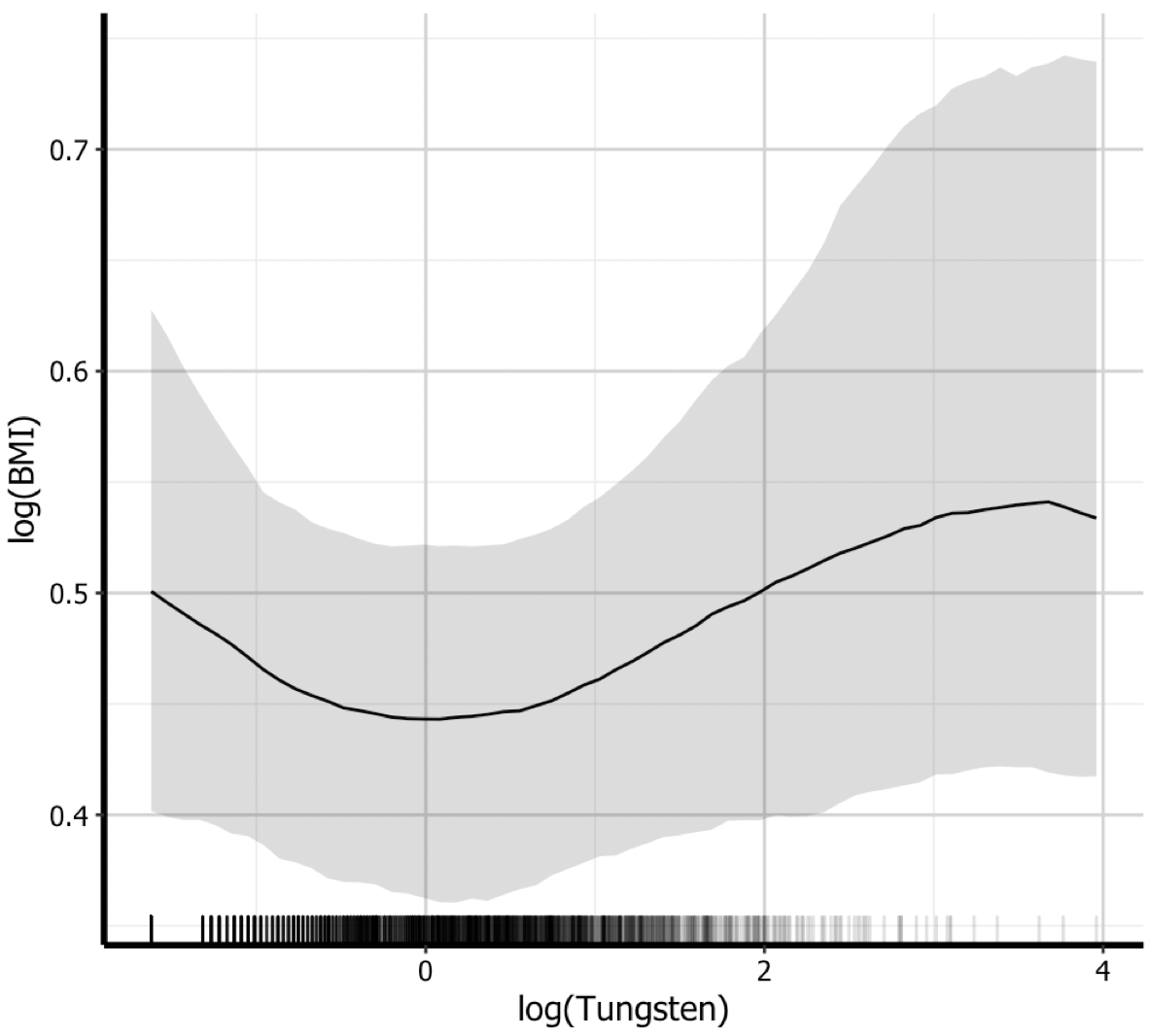}
\end{minipage}
\hfill
\end{figure}
\begin{figure}[H]
\vspace*{-2cm}
\begin{minipage}{0.48\textwidth}
\centering
\includegraphics[width=\linewidth, height= 5.8cm]{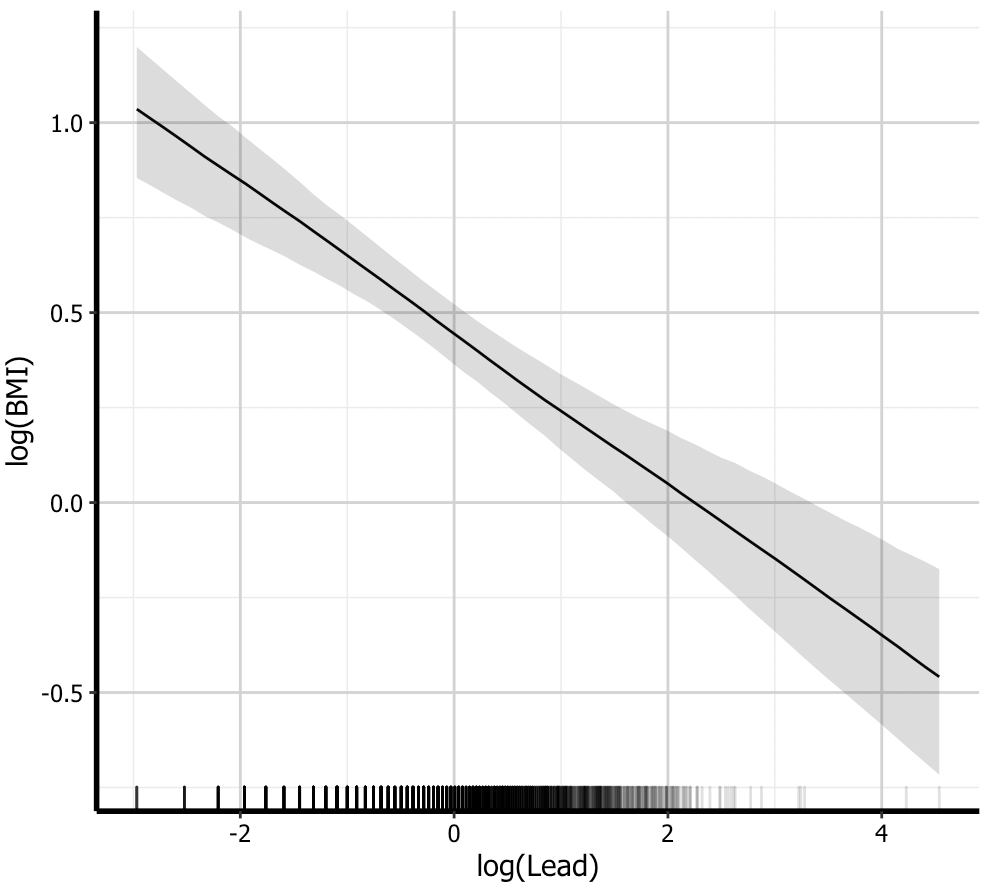}
\end{minipage} \hfill
\begin{minipage}{0.48\textwidth}
\centering
\hspace{0.1cm}\includegraphics[width=\linewidth, height= 5.8cm]{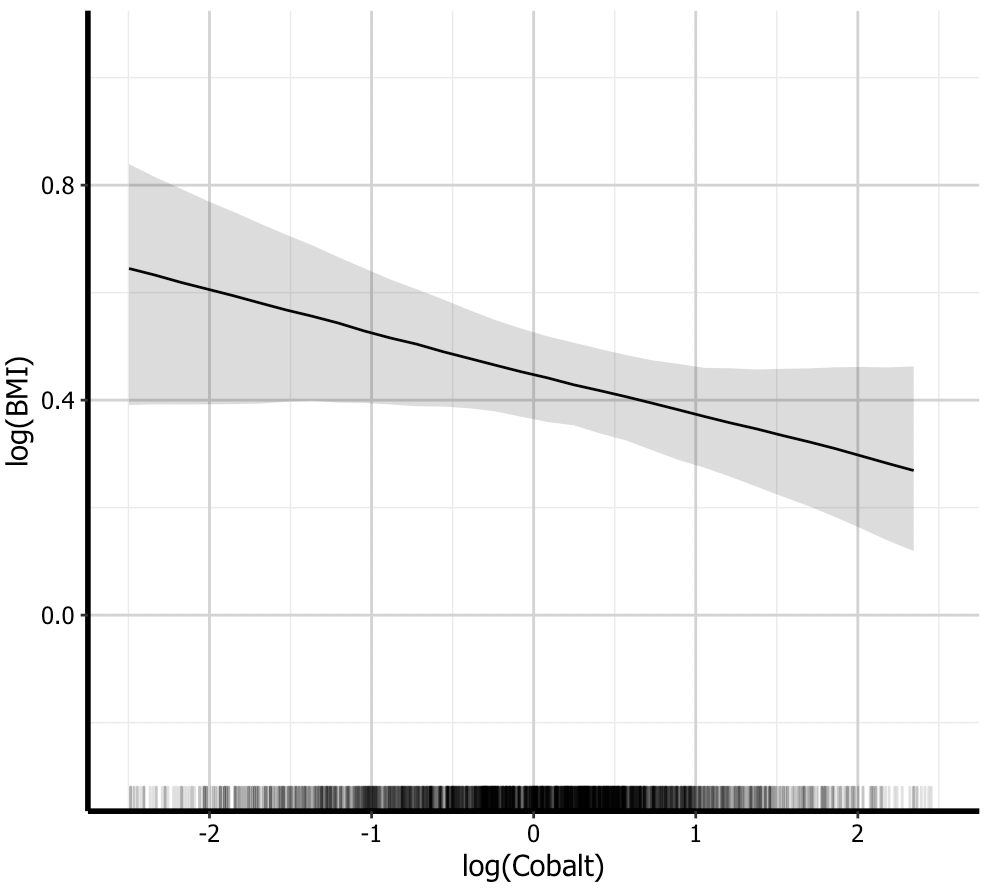}
\end{minipage}
\hfill
\caption{Estimated dose response curves for the chemicals Cadmium,  Tungsten, Lead and Cobalt, when all the other quantities are equal to their median. The black line corresponds to the posterior median, the shaded bands indicate 95\% posterior credible intervals, and the marks on the x-axis indicate the observed data points.}
\end{figure}

\textit{Figure 3} shows the estimated main effects of the chemicals, and $95\%$ credible intervals for MixSelect. Notice that most of the main effect estimates of the other models are equal to $0$, perhaps due to low power. The method PIE also estimates a negative association for Lead and Molybdenum; RAMP and hierNet estimate a negative association for Lead. Finally, there is suggestive evidence of a negative association between BMI with Cesium, Tin and Cobalt, which is also detected by PIE. In the \textit{\href{https://nbviewer.jupyter.org/github/fedfer/fedfer.github.io/blob/master/gp_AoAS_supplementary.pdf}{Supplementary Materials}} we consider possible chemical interactions with Sex and non-Hispanic Black ethnicity. The nonlinear effect of Cadmium in Females and non-Hispanic Blacks has a hill-shaped dose response as in \textit{Figure 2}, whereas it is negatively associated with BMI in the Male subgroup. Moreover, we found that Lead and Molybdenum exposures have a stronger negative effect on Females than Males, and we observe the opposite behavior for Tin and Cobalt.

\begin{table}[!htbp] \centering 
  \label{} 
\begin{tabular}{@{\extracolsep{5pt}} ccccccc} 
\\[-1.8ex]\hline 
\hline \\[-1.8ex] 
 & MixSelect & BKMR & hierNet & Family & PIE & RAMP \\ 
\hline \\[-1.8ex] 
in sample MSE & $0.530$ & $0.031$ & $0.573$ & $0.879$ & $0.626$ & $0.572$ \\ 
out of sample MSE & $0.687$ & $0.919$ & $0.611$ & $0.927$ & $0.710$ & $0.604$ \\ 
\hline \\[-1.8ex] 
\end{tabular} 

\caption{Performance of MixSelect, BKMR, RAMP, hierNet, Family and PIE for in sample mean squared error when training on the complete cases and out of sample mean squared error when holding out 500 data points.} 
\end{table} 

\begin{figure}[htbp]
\centering
\includegraphics[width=\linewidth, height= 6cm]{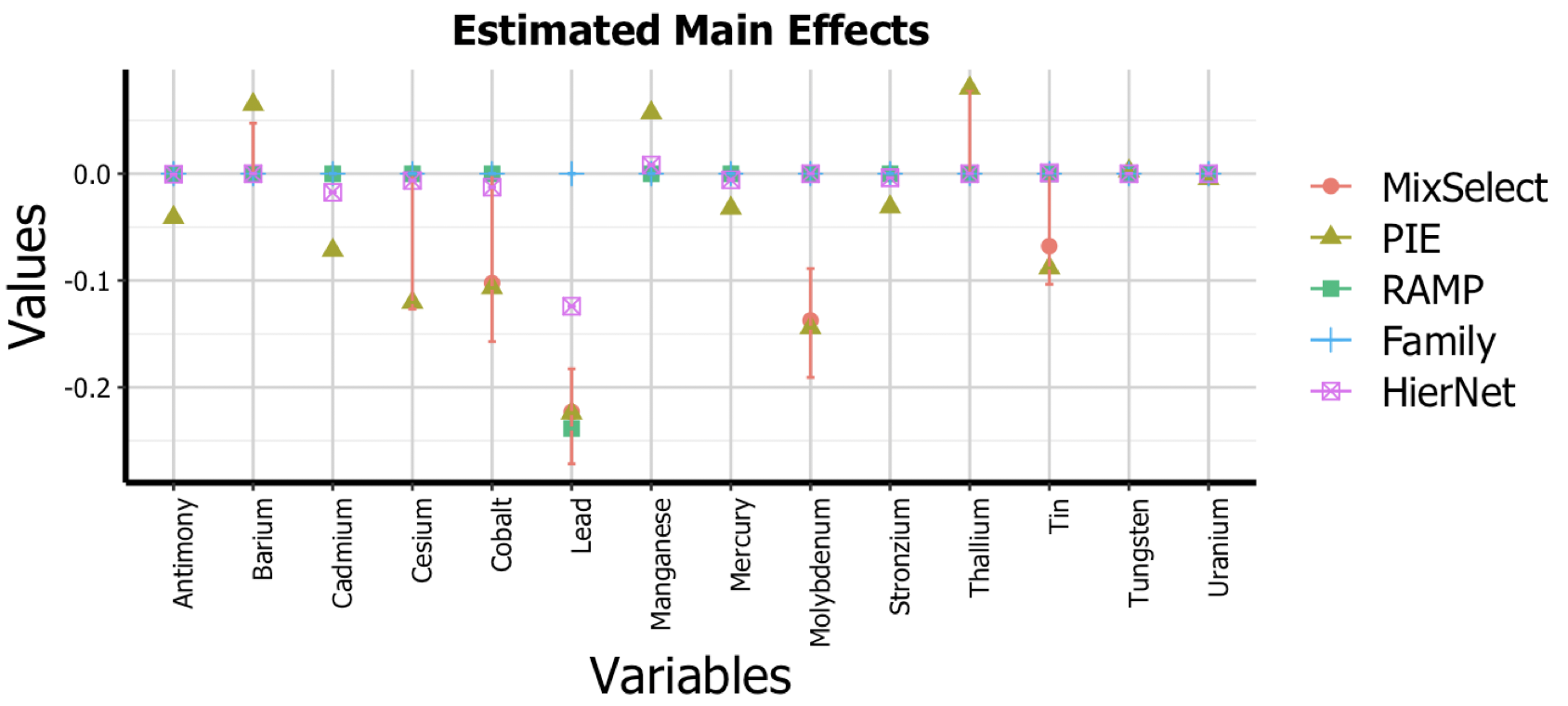}
\caption{Estimated main effects using MixSelect with $95\%$ credible intervals and estimated coefficients using RAMP, hierNet, Family and PIE. We trained all the methods on the dataset with complete cases. Exposure measurements are on the log scale.}
\end{figure}

\section{Discussion}

We proposed a MixSelect framework that allows identification of main effects and interactions. We also allow flexible nonlinear deviations from the parametric specification relying on a Gaussian process prior. We showed that MixSelect improves on the state-of-the-art for assessing associations between chemical exposures and health outcomes. To our knowledge, this is the first flexible method that is designed to provide interpretable estimates for main effects and interactions of chemical exposures while not constraining the model to have a simple parametric form. We also included variable selection, uncertainty quantification, missingness in the predictors, and limit of detection. The proposed specification provides a nice building block for more complicated data structures; for example, there are straightforward extensions to allow censored outcomes, longitudinal data, spatial dependence, and other issues.

NHANES data are obtained using a complex sampling design, that includes oversampling of certain population subgroups, and contains sampling weights for each observation that are inversely proportional to the probability of being sampled. We did not employ sampling weights in our analysis because our goal was to study the association between metals and BMI rather than providing population estimates. One possibility to include the sampling weights in our method is to jointly model the outcome and the survey weights \citep{si2015bayesian}, without assuming that the population distribution of strata is known. 

With correlated features, variable selection techniques can lead to multiple models having almost the same posterior probability of being the best one, and with few observations the interpretation of results becomes difficult. However, our method provided better inference under correlated predictors than BKMR \citep{BKMR2014}. We believe this is due to the projection approach, which protects against overfitting by adding a constraint to the highly flexible nonparametric surface. An alternative solution is to cluster the predictors at each iteration of the MCMC algorithm using a nonparametric prior specification for the coefficients \citep{maclehose2007bayesian}.

Instead of focusing on mean regression, we can easily modify MixSelect to accommodate quantile regression. In order to induce a regression on a specific quantile, one can use $(2.1)$ but with the residual $\epsilon_i$ having an asymmetric Laplace distribution \citep{yu2001bayesian}. The asymmetric Laplace can be represented as a scale mixture of Gaussians, facilitating a straightforward modification to our MCMC algorithm; refer to \cite{chen2013bayesian} for related work. Alternatively, it is possible to allow main effects and interactions to vary with quantiles of $y_i$, see for example \cite{reich2011bayesian}. We can also induce a quantile dependence on the nonlinear deviation $g^*(x_i)$. In particular, we can introduce uniformly distributed latent variables $\eta_i$ modifying the nonlinear deviation as $g^*(x_i, \eta_i)$, which is referred to as the Gaussian process transfer prior \citep{kundu2014latent}. 

Chemical studies usually involve up to dozens of exposures, but recent developments employing novel data collection techniques are starting to produce interesting datasets in which the number of exposures is in the order of the number of data points, so that the estimation of statistical interactions becomes infeasible with standard techniques. In this paper we impose heredity constraints and an approximation to the Gaussian process surface in order to deal with this problem, but new developments for dimension reduction are needed to scale up to allow massive number of exposures.

\section*{Acknowledgements}

This research was supported by grant 1R01ES028804-01 of the National Institute of Environmental Health Sciences of the United States Institutes of Health. The authors would like to thank Amy Herring and Alessandro Zito for helpful comments.

\begin{algorithm}[!htb]
\caption{MCMC algorithm for sampling the parameters of model (2.1)}
\begin{algorithmic} 
\vspace{0.5cm}
\STATE \textit{Step 1} Sample $\gamma_j$ for $j = 1,\cdots,p$ from
\begin{align*}
\pi(\gamma_j | \cdot)  \sim Bernoulli \big(\frac{1}{1 + \frac{1 -\pi}{\pi} R_j}\big)
\end{align*} 

\hspace*{1cm}  where $R_j = \frac{|X_{0j}^T \Sigma^{-1} X_{0j} + I|^{-1/2} exp(\frac{1}{2} m_0^T V_0 m_0)}{|X_{1j}^T \Sigma^{-1} X_{1j} + I|^{-1/2}exp(\frac{1}{2}m_1^T V_1 m_1)}$, $\Sigma = \sigma^2 I_n + PcP^T$, $m_0 = X_{0j}^T \Sigma^{-1} y$ \hspace*{1cm} and  $V_0 = (X_{0j}^T \Sigma^{-1} X_{0j}^T + I)^{-1}$. $X_{0j}$ is the matrix of covariates such that $\gamma_k = 1$ \hspace*{1cm} for  $k  \neq j$. $X_{1j}$  is the matrix of covariates such that $\gamma_k = 1$ for $k = 1,\cdots,p$, with \hspace*{1cm} $X_j$   included.
\vspace*{0.25cm}

\STATE \textit{Step 2} Sample $\pi$ from $\pi (\pi | \cdot) \sim Beta(a_\pi + \sum_{j=1}^p \gamma_j, b_\pi+ p - \sum_{j=1}^p \gamma_j)$
\vspace*{0.4cm}
\STATE \textit{Step 3} Sample the main coefficients $\beta_\gamma$ from the distribution: 
\begin{align*}
\pi (\beta_\gamma | \cdot) \sim N(V X_\gamma^T \Sigma^{-1}(y - \alpha Z - diag(X \Lambda X^T)), V)
\end{align*}

\hspace*{1cm}  where $V = (X_\gamma \Sigma^{-1} X_\gamma + I)^{-1}$ and the subscript $\gamma$ indicates that we are including \hspace*{1cm}  only the variables such that $\gamma_j = 1$
\vspace*{0.25cm}
\STATE \textit{Step 4} Set $\lambda_{j,k}$ equal to zero according to the chosen heredity condition. Then update \hspace*{1cm} $\lambda_{j,k }$ following an appropriate modification of \textit{Step 2}
\vspace*{0.25cm}
\STATE \textit{Step 5} Sample $\alpha$ following an appropriate modification of \textit{Step 2}
\vspace*{0.25cm}
\STATE \textit{Step 6} If $\gamma_\tau = 0$, set $\rho_j = 0$ and $\gamma_j^\rho = 0$ and move to \textit{Step 7}, else go to \textit{Step 6'}.  
\vspace*{0.25cm}
\STATE \textit{Step 6'} If $\rho_j \neq 0 $, perform \textit{delete} move: propose $\rho_j^* = 0 $ and $\gamma_j^* = 0 $. If $\rho_j = 0 $ perform \hspace*{1cm} \textit{add} move: propose $\rho_j^* > 0 $ and $\gamma_j^* = 1 $, for $j = 1,\cdots,p$. Compute $U_m^* D^* U_m^{* T}$ \hspace*{1cm} with the approximation of \textit{Section 3}, $\Sigma^{* -1}$ with Sherman-Woodbury formula \hspace*{1cm} and $|\Sigma^{* -1}|$ with determinant lemma. Then compute:
\begin{align*}
-2 \ \text{log}( r) = \text{log}|\Sigma^{* -1}| - \text{log}|\Sigma^{-1}| + \frac{1}{2}\mu^T (\Sigma^{* -1} - \Sigma^{ -1}) \mu,
\end{align*}

\hspace*{1cm} where $\mu = y - (Z \alpha + X \beta + diag(X \Lambda X^T) )$ . Sample $u$ from a Uniform distribution \hspace*{1cm} in the interval  $(0,1)$ and if $\text{log}(r) >\text{log}(u)  $, set $\rho_j = \rho_j^*$, $\gamma_j$, $\Sigma = \Sigma^*$, $|\Sigma^{ -1}|=$ \hspace*{1cm} $ = |\Sigma^{* -1}|$
\vspace*{0.25cm}
\STATE \textit{Step 7} For all $j = 1, \cdots, p$ such that $\rho_j \neq 0 $, perform a \textit{Gibbs-type} move: sample $\rho_j^*$ \hspace*{1cm} from a symmetric proposal distribution and then follow \textit{Step 5}.
\vspace*{0.25cm}
\STATE  \textit{Step 8} Sample $\varphi$ following an appropriate modification of \textit{Step 2}.
\vspace*{0.25cm}
\STATE  \textit{Step 9} Sample $\tau^{*2}$ from a symmetric proposal distribution and update following an \hspace*{1cm} appropriate modification of \textit{Step 5}. If $\tau^{*2} \neq 0$ perform a \textit{Gibbs-type} move.
\vspace*{0.25cm}
\STATE  \textit{Step 10}  $\pi(\sigma^{2} | \cdot) \sim InvGamma(\frac{1+n}{2}, \frac{1+ \mu^T (I_n + Pc'P^T)^{-1} \mu}{2})$ where $c'(x,x^*) = (\tau^{*})^2 \exp \bigg\{ \sum_{j=1}^p \rho_j (x_j-x^*_j)^2 \bigg\}$

\vspace{0.8cm}
\end{algorithmic}
\end{algorithm}

\bibliographystyle{plainnat}
\bibliography{ref}


\end{document}